\documentclass[11pt,journal,onecolumn]{IEEEtran}
\usepackage{amsfonts,amsmath,epsfig,graphicx,latexsym,mathrsfs,amssymb,marvosym,cite}
\title{Queue Length Asymptotics for Generalized Max-Weight Scheduling in the presence of Heavy-Tailed Traffic}
\author{\IEEEauthorblockN{Krishna Jagannathan, Mihalis Markakis, Eytan Modiano, and John N. Tsitsiklis\thanks{This work was supported by NSF grants CNS-0626781, CNS-0915988, and CCF-0728554, and by ARO Muri grant number W911NF-08-1-0238.}}
\IEEEauthorblockA{
Massachusetts Institute of Technology\\
Cambridge, MA 02139\\
Email: \{krishnaj, mihalis, modiano, jnt\}@mit.edu}
}

\newtheorem{corollary}{Corollary}
\newtheorem{lemma}{Lemma}
\newtheorem{proposition}{Proposition}
\newtheorem{remark}{Remark}
\newtheorem{theorem}{Theorem}
\newtheorem{definition}{Definition}

\newcommand{\bearno}{\begin{eqnarray*}}
\newcommand{\enarno}{\end{eqnarray*}}
\newcommand{\pr}[1]{{\mathbb P}\left\{#1\right\}}
\newcommand{\qed}{\hfill $\Box$}
\newcommand{\expect}[1]{{\mathbb E}\left[ #1 \right]}

\begin{document}

\maketitle

\begin{abstract}
We investigate the asymptotic behavior of the steady-state queue length distribution under generalized max-weight scheduling in the presence of heavy-tailed traffic. We consider a system consisting of two parallel queues, served by a single server. One of the queues receives heavy-tailed traffic, and the other receives light-tailed traffic. We study the class of throughput optimal max-weight-$\alpha$ scheduling policies, and derive an exact asymptotic characterization of the steady-state queue length distributions. In particular, we show that the tail of the light queue distribution is heavier than a power-law curve, whose tail coefficient we obtain explicitly. Our asymptotic characterization also contains an intuitively surprising result -- the celebrated max-weight scheduling policy leads to the \emph{worst possible} tail of the light queue distribution, among all non-idling policies.

Motivated by the above `negative' result regarding the max-weight-$\alpha$ policy, we analyze a log-max-weight (LMW) scheduling policy. We show that the LMW policy guarantees an exponentially decaying light queue tail, while still being throughput optimal.
\end{abstract}

\section{Introduction}
Traditionally, traffic in telecommunication networks has  been modeled using Poisson and Markov-modulated processes. These simple traffic models exhibit `local randomness', in the sense that much of the variability occurs in short time scales, and only an average behavior is perceived at longer time scales. With the spectacular growth of packet-switched networks such as the internet during the last couple of decades, these traditional traffic models have been shown to be inadequate. This is because the traffic in packetized data networks is intrinsically more `bursty', and exhibits correlations over longer time scales than can be modeled by any Markovian point process. Empirical evidence, such as the famous Bellcore study on self-similarity and long-range dependence in ethernet traffic \cite{leland} lead to increased interest in traffic models with high variability.

Heavy-tailed distributions, which have long been used to model high variability and risk in finance and insurance, were considered as viable candidates to model traffic in data networks. Further, theoretical work such as \cite{HRS96}, linking heavy-tails to long-range dependence (LRD) lent weight to the belief that extreme variability in the internet file sizes is ultimately responsible for the LRD traffic patterns reported in \cite{leland} and elsewhere.

Many of the early queueing theoretic results for heavy-tailed traffic were obtained for the single server queue; see \cite{PW00,boxma2007tails,BBNZ03} for surveys of these results. It turns out that the service discipline plays an important role in the latency experienced in a queue, when the traffic is heavy-tailed. For example, it was shown in \cite{anantharam} that any non-preemptive service discipline leads to infinite expected delay, when the traffic is sufficiently heavy-tailed. Further, the asymptotic behavior of latency under various service disciplines such as first-come-first-served (FCFS) and processor sharing (PS), is markedly different under light-tailed and heavy-tailed scenarios \cite{BBNZ03,WZ10}. This is important, for example, in the context of scheduling jobs in server farms \cite{HSY09}.

In the context of communication networks, a subset of the traffic flows may be well modeled as heavy-tailed, and the rest better modeled as light-tailed. In such a scenario, there are relatively few studies on the problem of scheduling \emph{between} the different flows, and the ensuing nature of interaction between the heavy-tailed and light-tailed traffic. Perhaps the earliest, and one of the most important studies in this category is \cite{BMv03}, where the interaction between light and heavy-tailed traffic flows under generalized processor sharing (GPS) is studied. In that paper, the authors derive the asymptotic workload behavior of the light-tailed flow, when its GPS weight is greater than its traffic intensity.

One of the key considerations in the design of a scheduling policy for a queueing network is \emph{throughput optimality}, which is the ability to support the largest set of traffic rates that is supportable by a given queueing network. Queue length based scheduling policies, such as max-weight scheduling \cite{TE93,TE92} and its many variants, are known to be throughput optimal in a general queueing network. For this reason, the max-weight family of scheduling policies has received much attention in various networking contexts, including switches \cite{mckeown1999achieving}, satellites \cite{NMR02}, wireless \cite{NMR03}, and optical networks \cite{brzezinski2005dynamic}.

In spite of a large and varied body of literature related to max-weight scheduling, it is somewhat surprising that the policy has not been adequately studied in the context of heavy-tailed traffic. Specifically, a question arises as to what behavior we can expect due to the interaction of heavy and light-tailed flows, when a throughput optimal max-weight-like scheduling policy is employed. Our present work is aimed at addressing this basic question.

In a recent paper \cite{MMT09}, a special case of the problem considered here is studied. Specifically, it was shown that when the heavy-tailed traffic has an infinite variance, the light-tailed traffic experiences an infinite expected delay under max-weight scheduling. Further, it was shown that the max-weight policy can be tweaked to favor the light-tailed traffic, so as to make the expected delay of the light-tailed traffic finite. In the present paper, we considerably generalize these results by providing a precise asymptotic characterization of the occupancy distributions under the max-weight scheduling family, for a large class of heavy-tailed traffic distributions.

We study a system consisting of two parallel queues, served by a single server. One of the queues is fed by a heavy-tailed arrival process, while the other is fed by light-tailed traffic. We refer to these queues as the `heavy' and `light' queues, respectively. In this setting, we analyze the asymptotic performance of max-weight-$\alpha$ scheduling, which is a generalized version of max-weight scheduling. Specifically, while max-weight scheduling makes scheduling decisions by comparing the queue lengths in the system, the max-weight-$\alpha$ policy uses different powers of the queue lengths to make scheduling decisions. Under this policy, \emph{we derive an exact asymptotic characterization of the light queue occupancy distribution}, and \emph{specify all the bounded moments of the queue lengths}.

A surprising outcome of our asymptotic characterization is that the `plain' max-weight scheduling policy induces the worst possible asymptotic behavior on the light queue tail. We also show that by a choice of parameters in the max-weight-$\alpha$ policy that increases the preference afforded to the light queue, the tail behavior of the light queue can be improved. Ultimately however, the tail of the light queue distribution is lower bounded by a power-law-like curve, for \emph{any} scheduling parameters used in the max-weight-$\alpha$ scheduling policy. Intuitively, the reason max-weight-$\alpha$ scheduling induces a power-law-like decay on the light queue distribution is that the light queue has to compete with a typically large heavy queue for service.

The simplest way to guarantee a good asymptotic behavior for the light queue distribution is to give the light queue complete priority over the heavy queue, so that it does not have to compete with the heavy queue for service. We show that under priority for the light queue, the tail distributions of \emph{both} queues are asymptotically as good as they can possibly be under any policy. Be that as it may, giving priority to the light queue has an important shortcoming -- it is not throughput optimal for a general constrained queueing system.

We therefore find ourselves in a situation where on the one hand, the throughput optimal max-weight-$\alpha$ scheduling leads to poor asymptotic performance for the light queue. On the other hand, giving priority to the light queue leads to good asymptotic behavior for both queues, but is not throughput optimal in general. To remedy this situation, we propose a throughput optimal log-max-weight (LMW) scheduling policy, which gives significantly more importance to the light queue compared to max-weight-$\alpha$ scheduling. We analyze the asymptotic behavior of the LMW policy and show that the\emph{ light queue occupancy distribution decays exponentially}. We also obtain the exact large deviation exponent of the light queue tail under a regularity assumption on the heavy-tailed input. Thus, the LMW policy has both desirable attributes -- it is throughput optimal, {and} ensures an exponentially decaying tail for the light queue distribution.

The remainder of this paper is organized as follows. In Section \ref{model}, we describe the system model. In Section \ref{math}, we present the relevant  definitions and mathematical preliminaries. Section \ref{Prioritysec} deals with the queue length behavior under priority scheduling. Sections \ref{mwalpha} and \ref{logmax} respectively contain our asymptotic results for max-weight-$\alpha$ scheduling, and the LMW policy. We conclude the paper in Section \ref{concl}.
\section{System Model}\label{model}

\begin{figure}
\begin{center}
\includegraphics[width=6cm]{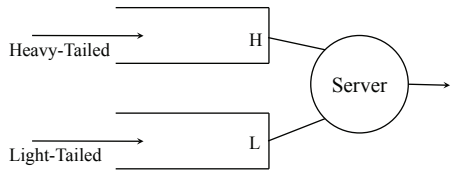}
\caption{A system of two parallel queues, with one of them fed with heavy-tailed traffic.}
\label{fig1}
\end{center}
\end{figure}
Our system consists of two parallel queues, $H$ and $L$, served by a single server, as depicted in Fig.~\ref{fig1}. Time is slotted, and stochastic arrivals of packet bursts occur to each queue in each slot. The server is capable of serving one packet per time slot from only one of the queues according to a scheduling policy.  Let $H(t)$ and $L(t)$ denote the number of packets that arrive during slot $t$ to $H$ and $L$ respectively. Although we postpone the precise assumptions on the traffic to Section \ref{arrival_assumptions}, let us loosely say that the input $L(t)$ is light-tailed, and $H(t)$ is heavy-tailed. We will refer to the queues $H$ and $L$ as the heavy and light queues, respectively. The queues are assumed to be always connected to the server. Let $q_H(t)$ and $q_L(t)$, respectively, denote the number of packets in $H$ and $L$ during slot $t,$ and let $q_H$ and $q_L$ denote the steady-state queue lengths, when they exist. Our aim is to characterize the behavior of $\pr{q_L>b}$ and $\pr{q_H>b}$ as $b$ becomes large, under various scheduling policies.

\section{Definitions and Mathematical Preliminaries}\label{math}
\subsection{Heavy-tailed distributions}
We begin by defining some properties of tail distributions of non-negative random variables.
\begin{definition}\label{light}
A random variable $X$ is said to be \emph{light-tailed} if there exists $\theta>0$ for which $\expect{\exp(\theta X)}<\infty.$ A random variable is \emph{heavy-tailed} if it is not light-tailed.
\end{definition}
In other words, a light-tailed random variable is one that has a well defined moment generating function in a neighborhood of the origin. The complementary distribution function of a light-tailed random variable decays at least exponentially fast. Heavy-tailed random variables are those which have complementary distribution functions that decay slower than any exponential. This class is often too general to study, so sub-classes of heavy-tailed distributions, such as sub-exponentials have been defined and studied in the past \cite{sigman}. We now review some definitions and existing results on some relevant classes of heavy-tailed distributions. In the remainder of this section, $X$ will denote a non-negative random variable, with complementary distribution function $\overline{F}(x)=\pr{X>x}.$ For the most part, we adhere to the terminology in \cite{BGT87,cline94}.

\emph{Notation:}  If $f(x)$ and $g(x)$ are positive functions defined on $[0,\infty],$ we write $f(x)\sim g(x)$ to mean $$\lim_{x\to\infty}\frac{f(x)}{g(x)}=1.$$ Similarly, $f(x)\gtrsim g(x)$ means  $$\liminf_{x\to\infty}\frac{f(x)}{g(x)}\geq1.$$
\begin{definition}
\label{reg}

\begin{enumerate}
\item  $\overline{F}(x)$ is said to have a  \emph{regularly varying} tail of index $\nu$, notation $\overline{F}\in\mathcal {R}(\nu),$ if $$\lim_{x\to\infty}\frac{\overline{F}(k x)}{\overline{F}(x)}=k^{-\nu},\ \forall\ k>0.$$ %$\overline{F}(x)$ is said to be \emph{regularly varying}, notation $\overline{F}\in\mathcal R$, if there exists $\eta>0$ such that $$\lim_{x\to\infty}\frac{\overline{F}(k x)}{\overline{F}(x)}=k^{-\eta},\ \forallk>0.$$
\item $\overline{F}(x)$ is said to be \emph{extended-regularly varying}, notation $\overline{F}\in\mathcal {ER}$, if for some real $c,d>0, $ and $\Gamma>1,$
$$k^d\le\liminf_{x\to\infty}\frac{\overline{F}(k x)}{\overline{F}(x)}\leq\limsup_{x\to\infty}\frac{\overline{F}(k x)}{\overline{F}(x)}\le k^c,\ \forall k\in[1,\Gamma].$$
\item $\overline{F}(x)$ is said to be \emph{intermediate-regularly varying}, notation $\overline{F}\in\mathcal {IR}$, if $$\lim_{k\downarrow1}\liminf_{x\to\infty}\frac{\overline{F}(k x)}{\overline{F}(x)}=\lim_{k\downarrow1}\limsup_{x\to\infty}\frac{\overline{F}(k x)}{\overline{F}(x)}=1.$$

\item $\overline{F}(x)$ is said to be \emph{order-regularly varying}, notation $\overline{F}\in\mathcal {OR}$, if for some $\Gamma>1,$ $$0<\liminf_{x\to\infty}\frac{\overline{F}(k x)}{\overline{F}(x)}\leq\limsup_{x\to\infty}\frac{\overline{F}(k x)}{\overline{F}(x)}<\infty,\ \forall k\in[1,\Gamma].$$
\end{enumerate}
\end{definition}

It is easy to see from the definitions that $\mathcal R \subset \mathcal {ER} \subset \mathcal {IR} \subset \mathcal {OR}.$ In fact, the containments are proper, as shown in \cite{cline94}. Intuitively, $\mathcal R$ is the class of distributions with tails that decay according to a power-law with parameter $\nu.$ Indeed, it can be shown that $$\overline{F}\in\mathcal R \iff  \overline{F}(x)=U(x)x^{-\nu},$$ where $U(x)$ is a \emph{slowly varying} function, i.e, a function that satisfies $U(kx)\sim U(x),\ \forall k>0.$ The other three classes are increasingly more general, but as we shall see, they all correspond to distributions that are asymptotically heavier than some power-law curve. In what follows, a statement such as $X\in\mathcal {IR}$ should be construed to mean $\pr{X>x}\in\mathcal{IR}.$

Next, we define the lower and upper orders of a distribution.
\begin{definition}
\label{order}

\begin{enumerate}
\item The \emph{lower order} of $\overline{F}(x)$ is defined by $$\xi(\overline{F})=\liminf_{x\to\infty}-\frac{\log \overline{F}(x)}{\log x}.$$

\item The \emph{upper order} of $\overline{F}(x)$ is defined by $$\rho(\overline{F})=\limsup_{x\to\infty}-\frac{\log \overline{F}(x)}{\log x}.$$
\end{enumerate}
\end{definition}

It can be shown that for regularly varying distributions, the upper and lower orders coincide with the index $\nu$. It also turns out that  both the orders are finite for the class $\mathcal{OR},$ as asserted below.

\begin{proposition}
\label{ORorder}
$\rho(\overline{F})<\infty$ for every $\overline{F}\in\mathcal{OR}.$
\end{proposition}

\emph{Proof:} Follows from Theorem 2.1.7 \& Proposition 2.2.5 in \cite{BGT87}.\qed

The following result, which is a consequence of  Proposition \ref{ORorder}, shows that every $\overline{F}\in\mathcal{OR}$ is asymptotically heavier than a power-law curve.
\begin{proposition}\label{ORheavy}
Let $\overline{F}\in\mathcal{OR}$. Then, for each $\rho>\rho(\overline{F}),$ we have $x^{-\rho}=o(\overline{F}(x))$ as $x\to\infty.$
\end{proposition}
\emph{Proof:} See Equation (2.4) in \cite{RS2008}.

Definitions \ref{reg} and \ref{order} deal with asymptotic tail probabilities of a random variable. Next, we introduce the notion of tail coefficient, which is a \emph{moment} property.
\begin{definition}
\label{TC}
The \emph{tail coefficient} of a random variable $X$ is defined by $$C_X=\sup\{c\ge0\ |\ \expect{X^c}<\infty\}.$$
\end{definition}
In other words, the tail coefficient is the threshold where the power moment of a random variable starts to blow up. Note that the tail coefficient of a light-tailed random variable is infinite. On the other hand, the tail coefficient of a heavy-tailed random variable may be infinite (e.g., log-normal) or finite (e.g., Pareto). The next result shows that the tail coefficient and order are, in fact, closely related parameters.

\begin{proposition}\footnote{The first author is grateful to Jayakrishnan Nair (Caltech) for suggesting a proof of  Proposition \ref{jk} via a personal communication.}
The tail coefficient of $X$ is equal to the lower order of $\overline{F}(x).$ %The following statements are equivalent

\label{jk}
\end{proposition}
\emph{Proof:} Suppose first that the lower order is infinite, so that for any $s>0,$ we can find an $x$ large enough such that $$-\frac{\log \pr{X>x}}{\log x}>s.$$ Thus, for large enough $x,$ we have $$\pr{X>x}<x^{-s},\ \forall\ s>0.$$ This implies $\expect{X^c}<\infty$ for all $c>0.$ Therefore, the tail coefficient of $X$ is also infinite.

Next suppose that $\xi(\overline{F})\in(0,\infty).$ We will show that (i) $\expect{X^c}<\infty$ for all $c<\xi(\overline{F}),$ and (ii)$\expect{X^c}=\infty$ for all $c>\xi(\overline{F}).$ To show (i), we argue as above that for large enough $x,$ we have  $\pr{X>x}<x^{-s},$ when $s<\xi(\overline{F}).$ Thus, $\expect{X^c}<\infty$ for all $c<\xi(\overline{F}).$ To show (ii), let us consider some $s$ such that $c>s>\xi(\overline{F}).$ By the definition of $\xi(\overline{F})$ there exists a sequence $\{x_i\}$ that increases to infinity as $i\to\infty,$ such that $$-\frac{\log\pr{X>x_i}}{\log x_i}\leq s,\ \forall\ i\iff\pr{X>x_i}\geq x^{-s},\ \forall\ i.$$ Therefore, $$\expect{X^c}=\int_0^\infty x^c{\rm d}F_X(x)\geq  \int_{x_i}^\infty x^c{\rm d}F_X(x)\geq x_i^c\pr{X>x_i}\geq x_i^cx_i^{-s},\ \forall\ i,$$ from which it follows that $\expect{X^c}=\infty.$ Therefore, the tail coefficient of $X$ is equal to $\xi(\overline{F})$.\qed

We emphasize that Proposition \ref{jk} holds for \emph{any} random variable, regardless of its regularity properties. Finally, we show that any distribution in $\mathcal{OR}$ necessarily has a finite tail coefficient.

\begin{proposition}
\label{ORfiniteTC}
If $X\in\mathcal{OR},$ then $X$ has a finite tail coefficient.
\end{proposition}

\emph{Proof:} From  Proposition \ref{ORorder}, the upper order is finite: $\rho(\overline{F})<\infty.$ Thus, the lower order $\xi(\overline{F})$ is also finite. But  Proposition \ref{jk} asserts that lower order is equal to the tail coefficient. \qed

\subsection{Assumptions on the arrival processes}\label{arrival_assumptions}
 We are now ready to state the precise assumptions on the arrivals processes.
\begin{enumerate}
\item The arrival processes $H(t)$ and $L(t)$ are independent of each other, and independent of the current state of the system.
\item  $H(t)$ is independent and identically distributed (i.i.d.) from slot-to-slot.
\item  $L(t)$ is i.i.d. from slot-to-slot.
\item $L(\cdot)$ is light-tailed with $\expect{L(t)}=\lambda_L.$

\item $H(\cdot)\in\mathcal{OR}$ with tail coefficient $C_H>1,$ and $\expect{H(t)}=\lambda_H.$

\end{enumerate}
We also assume that $\lambda_L+\lambda_H<1,$ so that the input rate does not overwhelm the service rate. Then, it can be shown that the system is stable\footnote{The notion of stability used here is the positive recurrence of the system occupancy Markov chain.} under any non-idling policy, and that the steady-state queue lengths $q_H$ and $q_L$ exist.

\subsection{Residual and Age distributions}

Here, we define the residual and age distributions for the heavy-tailed input process, which will be useful later. First, we note that $H(\cdot)$ necessarily has a non zero probability mass at zero, since $\lambda_H<1.$ Define $H_+$ as the strictly positive part of $H(\cdot).$ Specifically, $$\pr{H_+=m}=\frac{\pr{H(\cdot)=m}}{1-\pr{H(\cdot)=0}}, \ m=1,2,\dots.$$ Note that $H_+$ has tail coefficient equal to $C_H,$ and inherits any regularity property of $H(\cdot).$

Now consider a discrete-time renewal process with inter-renewal times distributed as $H_+.$ Let $H_R\in\{1,2,\dots\}$ denote the residual random variable, and $H_A\in\{0,1,\dots\}$ the age of the renewal process \cite{gal_book}.\footnote{We have defined the residual time and age such that if a renewal occurs at a particular time slot, the age at that time slot is zero, and the residual time is equal to the length of the upcoming renewal interval.} The joint distribution of the residual and the age can be derived using basic renewal theory:

\begin{equation}\label{joint}
\pr{H_R=k,H_A=l}=\frac{\pr{H_+=k+l}}{\expect{H_+}}, \ k\in\{1,2\dots\},\ l\in\{0,1,\dots\}.
\end{equation} The marginals of $H_R$ and $H_A$ can be derived from (\ref{joint}):

\begin{equation}\pr{H_R=k}=\frac{\pr{H_+\geq k}}{\expect{H_+}}, \ k\in\{1,2,\dots\}.\label{residue}\end{equation}

\begin{equation}\pr{H_A=k}=\frac{\pr{H_+> k}}{\expect{H_+}}, \ k\in\{0,1,\dots\}.\label{age}\end{equation}
Next, let us invoke a useful result from the literature.
\begin{lemma}\label{ERlemma}
If $H(\cdot)\in\mathcal{OR},$ then $H_R\in\mathcal{ER},$ and
\begin{equation}\label{sup}
\sup_n \frac{n\pr{H_+>n}}{\pr{H_R>n}}<\infty.
\end{equation}
A corresponding result also holds for the age $H_A.$
\end{lemma}
\emph{Proof:} See \cite[Lemma 4.2(i)]{cline94}.\qed

Using the above, we prove the important result that the residual distribution is \emph{one order heavier }than the original distribution.
\begin{proposition}
\label{oneorderheavier}
If $H(\cdot)\in\mathcal{OR}$ has tail coefficient equal to $C_H,$ then $H_R$ and $H_A$ have tail coefficient equal to $C_H-1.$
\end{proposition}
\emph{Proof:}  According to (\ref{sup}), we have, for all $a$ and some real $\chi,$ $$-\log\pr{H_R> a}\leq -\log a -\log\pr{H_+>a}+\chi.$$ Let us now consider the lower order of $H_R:$
$$\liminf_{a\to\infty}-\frac{\log\pr{H_R> a}}{\log a}\leq\liminf_{a\to\infty}\frac{ -\log a -\log\pr{H_+>a}+\chi}{\log a}=C_H-1.$$ In the last step above, we have used the tail coefficient of $H_+.$  Since the lower order of $H_R$ equals its tail coefficient (Lemma \ref{jk}), the above relation shows that the tail coefficient of $H_R$ is \emph{at most} $C_H-1$.

Next, to show the opposite inequality, let us consider the \emph{duration} random variable, defined as $$H_D=H_R+H_A.$$ Using the joint distribution (\ref{joint}), we can obtain the marginal of $H_D$ as $$\pr{H_D=k}=\frac{k\pr{H_+=k}}{\expect{H_+}},\ k\in\{1,2,\dots\}.$$ Thus, for any $\epsilon>0,$ the $C_H-1-\epsilon$ moment of $H_D$ is finite: $$\expect{H_D^{C_H-1-\epsilon}}=\sum_{k\geq1}\frac{k^{C_H-1-\epsilon}k\pr{H_+=k}}{\expect{H_+}}=\frac{\expect{H_+^{C_H-\epsilon}}}{\expect{H_+}}<\infty.$$ Since $H_R$ is stochastically dominated by $H_D,$ it is immediate that $\expect{H_R^{C_H-1-\epsilon}}<\infty.$ Therefore, the tail coefficient of $H_R$ is \emph{at least} $C_H-1,$ and the proposition is proved.
\qed

\section{The Priority Policies}\label{Prioritysec}
In this section, we study the two `extreme' scheduling policies, namely priority for $L$ and priority for $H$. Our analysis helps us arrive at the important conclusion that the tail of the heavy queue is asymptotically insensitive to the scheduling policy. In other words, there is not much we can do to improve or hurt the tail distribution of $H$ by the choice of a scheduling policy.
Further, we show that giving priority to the light queue ensures the best possible asymptotic decay for \emph{both} the queue length distributions.
\subsection{Priority for $H$}
In this policy, $H$ receives service whenever it is non-empty, and $L$ receives service only when $H$ is empty. It should be intuitively clear at the outset that this policy is bound to have undesirable impact on the light queue. The reason we analyze this policy is that it gives us a best case scenario for the heavy queue. 

Our first result shows that the steady-state heavy queue occupancy is one order heavier than its input distribution.
\begin{theorem}
\label{Hprior}
Under priority scheduling for $H$, the steady-state queue occupancy distribution of the heavy queue satisfies the following bounds.
\begin{enumerate}

\item For every $\epsilon>0,$ there exists a $\kappa_H(\epsilon)$ such that
\begin{equation}
\label{HpriorHub}
\pr{q_H>b}<\kappa_H(\epsilon)b^{-(C_H-1-\epsilon)},\ \forall\ b.
\end{equation}

\item
\begin{equation}\label{HpriorHlb}
\pr{q_H>b}\geq \lambda_H\pr{H_R>b},\ \forall\ b.
\end{equation}
\end{enumerate}
Further, $q_H$ is a heavy-tailed random variable with tail coefficient equal to $C_H-1.$ That is, for each $\epsilon>0,$ we have
\begin{equation}\label{Hub}\expect{q_H^{C_H-1-\epsilon}}<\infty,\end{equation} and
\begin{equation}\label{Hlb}
\expect{q_H^{C_H-1+\epsilon}}=\infty.
\end{equation}

\end{theorem}
\emph{Proof:} Equation (\ref{Hub}) can be shown using a straightforward Lyapunov argument, along the lines of \cite[Proposition 6]{MMT09}. Equation (\ref{HpriorHub}) follows from  (\ref{Hub}) and the Markov inequality.

Next, to show (\ref{HpriorHlb}), we consider a time instant $t$ at steady-state, and write $$\pr{q_H(t)>b}=\pr{q_H(t)>b|q_H(t)>0}\pr{q_H(t)>0}=\lambda_H\pr{q_H(t)>b|q_H(t)>0}.$$ We have used Little's law at steady-state to write $\pr{q_H(t)>0}=\lambda_H.$ Let us now lower bound the term $\pr{q_H(t)>b|q_H(t)>0}.$ Conditioned on $H$ being non-empty, denote by $\tilde{B}(t)$ the number of packets that belong to the burst in service that still remain in the queue at time $t$. Then, clearly, $q_H(t)\geq\tilde{B}(t),$ from which $\pr{q_H(t)>b|q_H(t)>0}\geq\pr{\tilde{B}(t)>b}.$ Now, since the $H$ queue receives service whenever it is non-empty, it is clear that the time spent at the head-of-line by a burst is \emph{equal} to its size. It can therefore be shown that in steady-state, $\tilde{B}(t)$ is distributed according to the residual variable $H_R.$ Thus, $\pr{q_H(t)>b|q_H(t)>0}\geq\pr{H_R>b},$ and (\ref{HpriorHlb}) follows. Finally, (\ref{Hlb}) follows from (\ref{HpriorHlb}) and Proposition \ref{oneorderheavier}. \qed

When the distribution of $H(\cdot)$ is regularly varying, the lower bound (\ref{HpriorHlb}) takes on a power-law form that agrees with the upper bound (\ref{HpriorHub}).
\begin{corollary}
\label{HpriorCor}
 If $H(\cdot)\in\mathcal{R}(C_H),$ then $$\pr{q_H>b}>U(b)b^{-(C_H-1)},\ \forall\ b,$$ where $U(\cdot)$ is some slowly varying function.
 \end{corollary}
%\emph{ Proof:} If $H(\cdot)$ is regularly varying with index $C_H,$ the residual $H_R$ is regularly varying with index $C_H-1.$
Since priority for $H$ affords the most favorable treatment to the heavy queue, it follows that the asymptotic behavior of $H$ can be no better than the above under \emph{any} policy.
\begin{proposition}
\label{Hnobetter}
Under any scheduling policy, $q_H$ is heavy-tailed with tail coefficient at most $C_H-1.$ That is, Equation (\ref{Hlb}) holds for all scheduling policies.
\end{proposition}
\emph{Proof:} The tail probability $\pr{q_H>b}$ under any other policy stochastically dominates the tail under priority for $H$. Therefore, the lower bounds (\ref{HpriorHlb}) and (\ref{Hlb}) would hold for all policies. \qed

Interestingly, under priority for $H$, the steady-state light queue occupancy $q_L$ is also heavy-tailed with the \emph{same} tail coefficient as $q_H.$ This should not be surprising, since the light queue has to wait for the entire heavy queue to clear, before it receives any service.
\begin{theorem}
\label{LunderH}
Under priority for $H$, $q_L$ is heavy-tailed with tail coefficient $C_H-1.$ Furthermore, the tail distribution $\pr{q_L>b}$ satisfies the following asymptotic bounds.
\begin{enumerate}

\item For every $\epsilon>0,$ there exists a $\kappa_L(\epsilon)$ such that
\begin{equation}
\label{HpriorLub}
\pr{q_L>b}<\kappa_L(\epsilon)b^{-(C_H-1-\epsilon)}.
\end{equation}

\item If $H(\cdot)\in\mathcal{OR},$ then
\begin{equation}\label{HpriorLlb}
\pr{q_L>b}\gtrsim \lambda_H\pr{H_A>\frac b{\lambda_L}}
\end{equation}
\end{enumerate}
\end{theorem}
\emph{Proof:} The upper bound (\ref{HpriorLub}) is a special case of Theorem \ref{ub_1} given in the next section. Let us show (\ref{HpriorLlb}). Notice first that the lower bound (\ref{HpriorLlb}) is asymptotic, unlike (\ref{HpriorHlb}) which is exact. As before, let us consider a time $t$ at steady-state, and write using Little's law $$\pr{q_L(t)>b}\geq\pr{q_L(t)>b|q_H(t)>0}\pr{q_H(t)>0}=\lambda_H\pr{q_L(t)>b|q_H(t)>0}.$$ Let us denote by $\tilde{A}(t)$ the number of slots that the current head-of-line burst has been in service. Clearly then, $L$ has not received any service in the interval $[t-\tilde{A}(t),t],$ and has kept all the arrivals that occurred during the interval. Thus, conditioned on $H$ being non-empty, $q_L(t)\geq\sum_{\sigma=t-\tilde{A}(t)}^t L(\sigma).$ Next, it can be seen that in steady-state, $\tilde{A}(t)$ is distributed as the age variable $H_A.$ Putting it all together, we can write \begin{equation}\label{inter_Hlb}\pr{q_L>b}\geq \lambda_H\pr{q_L(t)>b|q_H(t)>0}\geq \lambda_H\pr{\sum_{i=1}^{H_A}L(i)>b}.\end{equation} Next, since $H(\cdot)\in\mathcal{OR},$ Lemma \ref{ERlemma} implies that $H_A\in\mathcal{ER}\subset\mathcal{IR}.$ We can therefore invoke Lemma \ref{rand_sum} in the appendix to write \begin{equation}\label{simLb}\pr{\sum_{i=1}^{H_A}L(i)>b}\sim \pr{H_A>\frac b{\lambda_L}}.\end{equation} Finally, (\ref{HpriorLlb}) follows from (\ref{inter_Hlb}) and (\ref{simLb}). \qed

We note that if $H(\cdot)$ is regularly varying, the lower bound (\ref{HpriorLlb}) takes on a power-law form that matches the upper bound (\ref{HpriorLub}).

\subsection{Priority for $L$}\label{priorityL}

We now study the policy that serves $L$ whenever it is non-empty, and serves $H$ only if $L$ is empty. This policy affords the best possible treatment to $L$ and the worst possible treatment to $H$, among all non-idling policies. Under this policy, $L$ is completely oblivious to the presence of $H$, in the sense that it receives service whenever it has a packet to be served. Therefore, $L$ behaves like a discrete time G/D/1 queue, with light-tailed inputs. Classical large deviation bounds can be derived for such a queue; see \cite{bigqueues} for example.

Recall that since $L(\cdot)$ is light-tailed, the log moment generating function $$\Lambda_L(\theta)=\log\expect{e^{\theta L(\cdot)}}$$ exists for some $\theta>0.$ Define \begin{equation}\label{innate}E_L=\sup\{\theta|\Lambda_L(\theta)-\theta<0\}.\end{equation}
\begin{proposition}
\label{LunderL}
Under priority for $L$, $q_L$ satisfies the large deviation principle (LDP)
\begin{equation}
\label{LDP}
\lim_{b\to\infty}-\frac1b\log\pr{q_L>b}=E_L
\end{equation}
\end{proposition}
In words, the above proposition asserts that the tail of $q_L$ is asymptotically exponential, with rate function $E_L.$ We will refer to $E_L$ as the \emph{intrinsic exponent} of the light queue. An equivalent expression for the intrinsic exponent that is often used in the literature is
\begin{equation}\label{innate2} E_L=\inf_{a>0}\frac1a\Lambda_L^*(1+a),\end{equation} where $\Lambda_L^*(\cdot)$ is the Fenchel-Legendre transform \cite{bigqueues} of $\Lambda_L(\theta).$

 It is clear that the priority policy for $L$ gives the best possible asymptotic behavior for the light queue, and the worst possible treatment for the heavy queue. Surprisingly however, it turns out that the heavy queue tail under priority for $L$ is asymptotically as good as it is under priority for $H$.
\begin{proposition}
\label{HunderL}
Under priority for $L$, $q_H$ is heavy-tailed with tail coefficient $C_H-1.$
\end{proposition}
\emph{Proof:} This is a special case of Theorem \ref{ub_1}, given in the next section.\qed

The above result also implies that the tail coefficient of $H$ cannot be worse than $C_H-1$ under any other scheduling policy.
\begin{proposition}
\label{Hnoworse}
Under any non-idling scheduling policy, $q_H$ has a tail coefficient of at least $C_H-1.$ That is, Equation (\ref{Hub}) holds for all non-idling scheduling policies.
\end{proposition}
\emph{Proof:} The tail probability $\pr{q_H>b}$ under any other policy is stochastically dominated by the tail probability under priority for $L$. \qed

Propositions \ref{Hnobetter} and \ref{Hnoworse} together imply the insensitivity of the heavy queue's tail distribution to the scheduling policy. We state this important result in the following theorem.

\begin{theorem}
\label{Hinsensitive}
Under \emph{any} non-idling scheduling policy, $q_H$ is heavy-tailed with tail coefficient equal to $C_H-1.$ Further, $\pr{q_H>b}$ satisfies bounds of the form (\ref{HpriorHub}) and (\ref{HpriorHlb}) under all non-idling policies.
\end{theorem}
Therefore, it is not possible to either improve or hurt the heavy queue's asymptotic behavior, by the choice of a scheduling policy.

It is evident that the light queue has the best possible asymptotic behavior under priority for $L$. Although priority for $L$ is non-idling, and therefore throughput-optimal in this simple setting, we are ultimately interested in studying more sophisticated network models, where priority for $L$ may not be throughput optimal. We therefore analyze the asymptotic behavior of general throughput optimal policies belonging to the max-weight family.

\section{Queue Length Asymptotics for Max-Weight-$\alpha$ Scheduling}\label{mwalpha}
In this section, we analyze the asymptotic tail behavior of the light queue distribution under max-weight-$\alpha$ scheduling. %Note that the heavy queue tail is characterized by theorem \ref{Hinsensitive} for all non-idling policies, including the max-weight-$\alpha$ family.
For fixed parameters $\alpha_H>0$ and $\alpha_L>0,$ the max-weight-$\alpha$ policy operates as follows: During each time slot $t,$ perform the comparison $$q_L(t)^{\alpha_L}\gtreqless q_H(t)^{\alpha_H},$$ and serve one packet from the queue that wins the comparison. Ties can be broken arbitrarily, but we break them in favor of the light queue for the sake of definiteness. Note that $\alpha_L=\alpha_H$ corresponds to the usual max-weight policy, which serves the longest queue in each slot. $\alpha_L/\alpha_H>1$ corresponds to emphasizing the light queue over the heavy queue, and vice-versa.

We provide an asymptotic characterization of the light queue occupancy distribution under max-weight-$\alpha$ scheduling by deriving matching upper and lower bounds. Our characterization shows that the light queue occupancy is heavy-tailed under max-weight-$\alpha$ scheduling for all values of the parameters $\alpha_H$ and $\alpha_L.$ %However, the tail coefficient of the light queue distribution can be improved if the $\alpha$ parameters are chosen to favor the light queue.
Since we obtain distributional bounds on the light queue occupancy, our results also shed further light on the moment results derived in \cite{MMT09} for max-weight-$\alpha$ scheduling.

\subsection{Upper bound}
In this section, we derive two different upper bounds on the overflow probability $\pr{q_L>b},$ that both hold under max-weight-$\alpha$ scheduling. However, depending on the values of $\alpha_H$ and $\alpha_L,$ one of them would be tighter. The first upper bound holds for all non-idling policies, including max weight-$\alpha$ scheduling.
\begin{theorem} \label{ub_1} Under \emph{any} non-idling policy, and for every $\epsilon>0,$
there exists a constant $\kappa_1(\epsilon)>0,$ such that
\begin{equation}\expect{q_L^{C_H-1-\epsilon}}<\infty\label{expec_1}\end{equation} and
\begin{equation}\label{tailub_1}\pr{q_L>b}<\kappa_1(\epsilon) b^{-(C_H-1-\epsilon)}.\end{equation}
\end{theorem}
\emph{Proof:} Let us combine the two queues into one, and consider
the sum input process $H(t)+L(t)$ feeding the composite queue. The
server serves one packet from the composite queue in each slot. Under
any non-idling policy in the original system, the occupancy of the composite queue is given by $q=q_H+q_L.$  Let us first show that the combined input has tail coefficient equal to $C_H.$

\begin{lemma}
The tail coefficient of $H(\cdot)+L(\cdot)$ is $C_H.$
\end{lemma}
\emph{Proof:} Clearly,
$\expect{(H+L)^{C_H+\delta}}\ge\expect{H^{C_H+\delta}}=\infty,$ for
every $\delta>0.$ We next need to show that
$\expect{(H+L)^{C_H-\delta}}<\infty,$ for
every $\delta>0.$ For a random variable $X$ and
event $E,$ let us introduce the notation
$\expect{X;E}=\expect{X1_E},$ where $1_E$ is the indicator of $E.$
(Thus, for example, $\expect{X}=\expect{X;E}+\expect{X;E^c}.$) Now,
\begin{eqnarray*}\expect{(H+L)^{C_H-\delta}}&=&\expect{(H+L)^{C_H-\delta};H>L}+\expect{(H+L)^{C_H-\delta};H\leq
L}\\&\leq&\expect{(2H)^{C_H-\delta};H>L}+\expect{(2L)^{C_H-\delta};H\leq
L}\\&<&2^{C_H-\delta}\left\{\expect{H^{C_H-\delta}}+\expect{L^{C_H-\delta}}\right\}<\infty
\end{eqnarray*}
where the last inequality follows from the tail coefficient of $H(\cdot),$ and the light-tailed nature of $L(\cdot).$\qed

The composite queue is therefore a G/D/1 queue with input tail coefficient $C_H.$
For such a queue, it can be shown that
\begin{equation}\label{comb_queue}\expect{q^{C_H-1-\epsilon}}<\infty.\end{equation} This is, in fact, a direct consequence of Theorem~\ref{Hprior}.

Thus, in terms of the queue lengths in the original system, we have
$$\expect{(q_H+q_L)^{C_H-1-\epsilon}}<\infty,$$ from which it is
immediate that $\expect{q_L^{C_H-1-\epsilon}}<\infty.$ This proves
(\ref{expec_1}). To show (\ref{tailub_1}), we use the Markov
inequality to write
$$\pr{q_L>b}=\pr{q_L^{C_H-1-\epsilon}>b^{C_H-1-\epsilon}}<\frac{\expect{q_L^{C_H-1-\epsilon}}}{b^{C_H-1-\epsilon}}<\kappa_1(\epsilon)
b^{-(C_H-1-\epsilon)}.$$ \qed

The above result asserts that the tail coefficient of $q_L$ is at least $C_H-1$ under any non-idling policy, and that $\pr{q_L>b}$ is uniformly upper bounded by a power-law curve. Our second upper bound is specific to max-weight-$\alpha$ scheduling. It hinges on a simple observation regarding the scaling of the $\alpha$ parameters, in addition to a theorem in \cite{MMT09}. We first state the following elementary observation due to its usefulness.\\
%\begin{lemma}\label{scaling}
\emph{Observation:} (Scaling of $\alpha$ parameters) Let $\alpha_H$ and $\alpha_L$ be given parameters of a max-weight-$\alpha$ policy, and let $\beta>0$ be arbitrary. Then, the max-weight-$\alpha$ policy that uses the parameters $\beta\alpha_H$ and $\beta\alpha_L$ for the queues $H$ and $L$ respectively, is \emph{identical} to the original policy. That is, in each time slot, the two policies make the same scheduling decision.
%\end{lemma}

Next, let us invoke an important result from \cite{MMT09}.
\begin{theorem}\label{mihalis}
If max-weight-$\alpha$ scheduling is performed with $0<\alpha_H< C_H-1,$
then, for \emph{any} $\alpha_L>0,$ we have $\expect{q_L^{\alpha_L}}<\infty.$
\end{theorem}
Thus, by choosing a large enough $\alpha_L,$ any moment of the light queue length can be made finite, as long as $\alpha_H<C_H-1$. Our second upper bound, which we state next, holds regardless of how the $\alpha$ parameters are chosen.
\begin{theorem}
Define $$\gamma=\frac{\alpha_L}{\alpha_H}(C_H-1).$$ Under max weight-$\alpha$ scheduling, and for every $\epsilon>0,$ there exists a constant $\kappa_2(\epsilon)>0,$ such that
\begin{equation}\expect{q_L^{\gamma-\epsilon}}<\infty\label{expec_2}\end{equation} and
\begin{equation}\label{tailub_2}\pr{q_L>b}<\kappa_2(\epsilon) b^{-(\gamma-\epsilon)}.\end{equation}
\label{ub_2}
\end{theorem}
\emph{Proof:} Given $\epsilon>0,$ let us choose $\beta=(C_H-1)/\alpha_H-\epsilon/\alpha_L,$ and perform max-weight-$\alpha$ scheduling with parameters $\beta\alpha_H$ and $\beta\alpha_L.$ According to the above observation, this policy is identical to the original max-weight-$\alpha$ policy. Next, since $\beta\alpha_H<C_H-1,$ Theorem~\ref{mihalis} applies, and we have $\expect{q_L^{\beta\alpha_L}}=\expect{q_L^{\gamma-\epsilon}}<\infty,$ which proves (\ref{expec_2}). Finally, (\ref{tailub_2}) can be proved using (\ref{expec_2}) and the Markov inequality.  \qed

The above theorem asserts that the tail coefficient of $q_L$ is at least $\gamma$ under the max weight-$\alpha$ policy.
We remark that Theorem~\ref{ub_1} and Theorem~\ref{ub_2} both hold for max-weight-$\alpha$ scheduling with any parameters. However, one of them yields a stronger bound than the other, depending on the $\alpha$ parameters. Specifically, we have the following two cases:

\begin{itemize}
\item[(i)] $\frac{\alpha_L}{\alpha_H}\leq1:$ This is the regime where the light queue is given lesser priority, when compared to the heavy queue. In this case, Theorem~\ref{ub_1} yields a stronger bound.

\item[(ii)] $\frac{\alpha_L}{\alpha_H}>1:$ This is the regime where the light queue is given more priority
compared to the heavy queue. In this case, Theorem~\ref{ub_2} gives the stronger bound.

\end{itemize}

\begin{remark} The upper bounds in this section hold whenever $H(\cdot)$ is heavy-tailed with tail coefficient $C_H.$ We need the assumption $H(\cdot)\in\mathcal{OR}$ only to derive the lower bounds in the next subsection.

\end{remark}

\subsection{Lower bound}
In this section, we state our main lower bound result, which asymptotically lower bounds the tail of the light queue distribution in terms of the tail of the residual variable $H_R.$
\begin{theorem}\label{main_lb}
Let $H(\cdot)\in\mathcal{OR}.$ Then, under max-weight-$\alpha$ scheduling with parameters $\alpha_H$ and $\alpha_L$, the distribution of the
light queue occupancy satisfies the following asymptotic lower bounds:
\begin{enumerate}
\item If $\frac{\alpha_L}{\alpha_H}<1,$ \begin{equation}\label{lb<} \pr{q_L\ge b}\gtrsim \lambda_H\pr{H_R\ge\frac{b}{\lambda_L}}\end{equation}
\item If $\frac{\alpha_L}{\alpha_H}=1,$ \begin{equation}\label{lb=} \pr{q_L\ge b}\gtrsim \lambda_H\pr{H_R\ge b\left(1+\frac{1}{\lambda_L}\right)}\end{equation}
\item If $\frac{\alpha_L}{\alpha_H}>1,$ \begin{equation}\label{lb>} \pr{q_L\ge b}\gtrsim \lambda_H\pr{H_R\ge b^{\alpha_L/\alpha_H}}.\end{equation}
\end{enumerate}
\end{theorem}
As a special case of the above theorem, when $H(\cdot)$ is regularly varying with index $C_H,$ the lower bounds take on a more pleasing power-law form that matches the upper bounds (\ref{tailub_1}) and (\ref{tailub_2}).
\begin{corollary}
Suppose $H(\cdot)\in\mathcal R (C_H).$ Then, under max-weight-$\alpha$ scheduling with parameters $\alpha_H$ and $\alpha_L$, the distribution of the light queue satisfies the following asymptotic lower bounds:
\begin{enumerate}
\item If $\frac{\alpha_L}{\alpha_H}\leq1,$ \begin{equation}\label{lbr<} \pr{q_L\ge b}\gtrsim U(b)b^{-(C_H-1)}\end{equation}
%\item If $\frac{\alpha_L}{\alpha_H}=1,$ \begin{equation}\label{lb=} \pr{q_L\ge b}\gtrsim \pr{H_R\ge b\left(1+\frac{1}{\lambda_L}\right)}\end{equation}
\item If $\frac{\alpha_L}{\alpha_H}>1,$ \begin{equation}\label{lbr>} \pr{q_L\ge b}\gtrsim U(b)b^{-\gamma},\end{equation}
\end{enumerate} where $U(\cdot)$ is some slowly varying function.
\end{corollary}

It takes several steps to prove Theorem~\ref{main_lb}; we start by defining and studying a related fictitious queueing system.

\subsection{Fictitious system}
The fictitious system consists of two queues, fed by the \emph{same input processes} that feed the original system. In the fictitious system, let us call the
queues fed by heavy and light traffic $\tilde{H}$ and $\tilde{L}$
respectively. The fictitious system operates under the
following service discipline.

\emph{Service for the fictitious system:} The queue $\tilde{H}$ receives
service in every time slot. The queue $\tilde{L}$ receives service
at time $t$ if and only if $q_{\tilde{L}}(t)^{\alpha_L}\geq q_{\tilde{H}}(t)^{\alpha_H}.$

Note that if $\tilde{L}$ receives service and $\tilde{H}$ is non-empty, \emph{two} packets are
served from the fictitious system. Also, $\tilde{H}$ is just a
discrete time $G/D/1$ queue, since it receives service at every time
slot. We now show a simple result which asserts that the light queue in
the original system is `longer' than in the fictitious system.

\begin{proposition}
\label{fict} Suppose a given input sample path feeds the queues in
both the original and the fictitious systems. Then, for all $t,$ it
holds that $q_{\tilde{L}}(t)\leq q_L(t).$ In particular, for each
$b>0,$ we have \[\pr{q_L>b}\geq \pr{q_{\tilde{L}}>b}.\]
\end{proposition}
\emph{Proof:} We will assume the contrary and arrive at a contradiction. Suppose $q_{\tilde{L}}(0)=q_L(0),$ and that for some time $t>0,$
$q_{\tilde{L}}(t)>q_L(t).$ Let $\tau>0$ be the first time when $q_{\tilde{L}}(\tau)>q_L(\tau).$ It is then necessary that $q_{\tilde{L}}(\tau-1)=q_L(\tau-1),$ since no more than one packet is served from a queue in each slot. Next, $q_{\tilde{L}}(\tau-1)=q_L(\tau-1),$ and $q_{\tilde{L}}(\tau)>q_L(\tau)$ together imply that $L$ received service at time $\tau-1,$ but $\tilde{L}$ did not. This is possible only if $q_H(\tau-1)<q_{\tilde{H}}(\tau-1),$ which is a contradiction, since $\tilde{H}$ receives service in each slot. \qed

Next, we show that the distribution of $q_{\tilde{L}}$ satisfies the
lower bounds in Equations (\ref{lb<})-(\ref{lb>}). Theorem~\ref{main_lb}  then follows, in light of  Proposition~\ref{fict}.

\begin{theorem}

\label{fictlb}
In the fictitious system, the distribution of $q_{\tilde{L}}$ is asymptotically lower bounded as follows.
\begin{enumerate}
\item If $\frac{\alpha_L}{\alpha_H}<1,$ \begin{equation}\label{lb<t} \pr{q_{\tilde{L}}>b}\gtrsim \lambda_H\pr{H_R>\frac{b}{\lambda_L}}\end{equation}
\item If $\frac{\alpha_L}{\alpha_H}=1,$ \begin{equation}\label{lb=t} \pr{q_{\tilde{L}}>b}\gtrsim \lambda_H\pr{H_R>b\left(1+\frac{1}{\lambda_L}\right)}\end{equation}
\item If $\frac{\alpha_L}{\alpha_H}>1,$ \begin{equation}\label{lb>t} \pr{q_{\tilde{L}}>b}\gtrsim \lambda_H\pr{H_R>b^{\alpha_L/\alpha_H}}\end{equation}
\end{enumerate}

\end{theorem}
\emph{Proof:} Let us consider an instant $t$ when the fictitious system is in steady-state. Since the heavy queue in the fictitious system receives service in each slot, the steady-state probability $\pr{q_{\tilde{H}}>0}=\lambda_H$ by Little's law. Therefore, we have the lower bound $$\pr{q_{\tilde{L}}>b}\geq \lambda_H\pr{q_{\tilde{L}}>b|q_{\tilde{H}}>0}.$$ In the rest of the proof, we will lower bound the above conditional probability.

Indeed, conditioned on $q_{\tilde{H}}>0,$ denote as before by $\tilde{B}(t),$ the number of packets that belong to the head-of-line burst that still remain in $\tilde{H}$ at time $t.$ Similarly, denote by $\tilde{A}(t)$ the number of packets from the head-of-line burst that have already been served by time $t.$ Since $\tilde{H}$ is served in every time slot, $\tilde{A}(t)$ also denotes the number of time
slots that the HoL burst has been in service at $\tilde{H}.$

The reminder of our proof shows that $q_{\tilde{L}}(t)$ stochastically dominates a particular heavy-tailed random variable. Indeed, at the instant $t,$ there are two possibilities:

\begin{itemize}

\item[(a)] $q_{\tilde{L}}(t)^{\alpha_L}\geq \tilde{B}(t)^{\alpha_H},$ or

\item[(b)] $q_{\tilde{L}}(t)^{\alpha_L}< \tilde{B}(t)^{\alpha_H}$,
\end{itemize}
Let us take a closer look at case (b) in the following proposition.

\begin{proposition}\label{caseb} Suppose that
$$q_{\tilde{L}}(t)^{\alpha_L}< \tilde{B}(t)^{\alpha_H}.$$ Let $\sigma\leq t$ be the instant before $t$ that $\tilde{L}$ last received service. Then, the current head-of-line burst arrived at $\tilde{H}$ \emph{after} the instant $\sigma.$
\end{proposition}
\emph{Proof:} We have $$q_{\tilde{H}}(\sigma)^{\alpha_H}\leq q_{\tilde{L}}(\sigma)^{\alpha_L}\leq
q_{\tilde{L}}(t)^{\alpha_L}<\tilde{B}(t)^{\alpha_H}.$$ The first inequality holds because $\tilde{L}$ received service at $\sigma,$ the second inequality is true since $\tilde{L}$ does not receive service between $\sigma$ and $t,$ and the final inequality is from the hypothesis.

We have shown that $q_{\tilde{H}}(\sigma)<\tilde{B}(t),$ and hence the HoL burst could not have arrived by the time slot $\sigma.$\qed

The above proposition implies that if case (b) holds, $\tilde{L}$ has not received service ever since the HoL burst arrived at $\tilde{H}$. In particular, $\tilde{L}$ has not received service for $\tilde{A}(t)$ time slots, and it accumulates all arrivals that occur during the interval $[t-\tilde{A}(t),t]$. Let us denote the number of arrivals to $\tilde{L}$ during this interval as $$S_{\tilde{A}}=\sum_{i=t-\tilde{A}(t)}^{t}L(i).$$ In this notation, our argument above implies that if case (b) holds, then $q_{\tilde{L}}(t)\geq S_{\tilde{A}}.$ Putting this together with case (a), we can conclude that

\begin{equation}\label{min}
q_{\tilde{L}}(t)\geq \min(\tilde{B}(t)^{\alpha_H/\alpha_L},\ S_{\tilde{A}}).
\end{equation}
Therefore,
\begin{equation}\label{minbnd1}
\pr{q_{\tilde{L}}(t)>b}\geq \lambda_H\pr{\tilde{B}(t)^{\alpha_H/\alpha_L}>b,\ S_{\tilde{A}}>b}.
\end{equation} Recall now that in steady-state, $\tilde{B}(t)$ is distributed as $H_R,$ and $\tilde{A}(t)$ is distributed as $H_A.$ Therefore, the above bound can be written as
\begin{equation}\label{minbnd2}
\pr{q_{\tilde{L}}>b}\geq \lambda_H\pr{H_R^{\alpha_H/\alpha_L}>b,\ \sum_{i=1}^{H_A} L(i)>b}.
\end{equation}
Lemma~\ref{mainlemma} shows that
$$\pr{H_R^{\alpha_H/\alpha_L}>b,\ \sum_{i=1}^{H_A} L(i)>b}\sim\left\{\begin{array}{cc} \pr{H_R\ge \frac{b}{\lambda_L}}&\frac{\alpha_L}{\alpha_H}<1,\\ \pr{H_R\ge b+\frac{b}{\lambda_L}}&\frac{\alpha_L}{\alpha_H}=1,\\ \pr{H_R\ge b^{\alpha_L/\alpha_H}}&\frac{\alpha_L}{\alpha_H}>1.\end{array}\right.$$ Notice that the assumption $H(\cdot)\in\mathcal{OR}$ is used in the proof of Lemma~\ref{mainlemma}.

Theorem~\ref{fictlb} now follows from the above asymptotic relation and (\ref{minbnd2}).\qed\\
\emph{Proof of Theorem~\ref{main_lb}:} The result follows from Theorem~\ref{fictlb} and  Proposition~\ref{fict}.\qed
\section{Tail Coefficient of $q_L$}

In this section, we characterize the exact tail coefficient of the light queue distribution under max-weight-$\alpha$ scheduling. In particular, we show that the upper bound (\ref{expec_1}) is tight for $\frac{\alpha_L}{\alpha_H}\le1,$ and (\ref{expec_2}) is tight for $\frac{\alpha_L}{\alpha_H}>1.$

\begin{theorem}
\label{tailcoeff}
The tail coefficient of the steady-state queue length $q_L$ of the light queue is given by
\begin{itemize}
\item[(i)] $C_H-1$ for $\frac{\alpha_L}{\alpha_H}\le1,$ and
\item[(ii)] $\gamma=\frac{\alpha_L}{\alpha_H}(C_H-1)$ for $\frac{\alpha_L}{\alpha_H}>1.$
\end{itemize}
\end{theorem}
\emph{Proof:} Consider first the case $\frac{\alpha_L}{\alpha_H}\le1.$ The lower order (Definition~\ref{order}) of $q_L$ can be upper bounded using (\ref{lb<}) or (\ref{lb=}) as follows
\begin{multline*}\liminf_{b\to\infty}-\frac{\log\pr{q_L>b}}{\log b}\le \liminf_{b\to\infty}-\frac{\log\lambda_H+\log\pr{H_R\geq \frac b{\lambda_L}}}{\log b}\\=\liminf_{a\to\infty}-\frac{\log\pr{H_R\geq a}}{\log a}=C_H-1.\end{multline*}
The last step is from  Proposition~\ref{oneorderheavier}. The above equation shows that the tail coefficient of $q_L$ is at most $C_H-1.$ However, it is evident from (\ref{expec_1}) that the tail coefficient of $q_L$ is \emph{at least} $C_H-1.$ Therefore, the tail coefficient of $q_L$ equals $C_H-1$ for $\frac{\alpha_L}{\alpha_H}\le1.$ This proves case (i) of the theorem.

Next, consider $\frac{\alpha_L}{\alpha_H}>1.$ Using (\ref{lb>}), we can upper bound the lower order of $q_L$ as
\begin{eqnarray}\liminf_{b\to\infty}-\frac{\log\pr{q_L>b}}{\log b}&\le&\liminf_{b\to\infty}-\frac{\log\pr{H_R\ge b^{\alpha_L/\alpha_H}}}{\log b}\nonumber\\ &=&\frac{\alpha_L}{\alpha_H}\liminf_{a\to\infty}\frac{-\log\pr{H_R\ge a}}{\log a}=\frac{\alpha_L}{\alpha_H}(C_H-1)\label{gam}\end{eqnarray}
Equation (\ref{gam}) shows that the tail coefficient of $q_L$ is at most $\gamma.$ However, it is  evident from (\ref{expec_2}) that the tail coefficient of $q_L$ is \emph{at least} $\gamma.$ Therefore, the tail coefficient of $q_L$ equals $\gamma=\frac{\alpha_L}{\alpha_H}(C_H-1)$ for $\frac{\alpha_L}{\alpha_H}>1.$ This proves case (ii) of the theorem.\qed

%Case (i) of the theorem asserts that the tail coefficient of the light queue does not change between max weight scheduling ($\alpha_L/\alpha_H=1$) and priority for $H$  ($\alpha_L/\alpha_H\downarrow0$). However, for $\alpha_L/\alpha_H>1,$ the tail coefficient of $q_L$ increases in proportion to the ratio of the $\alpha$ parameters.
\begin{figure}
\begin{center}
\includegraphics[width=8.5cm]{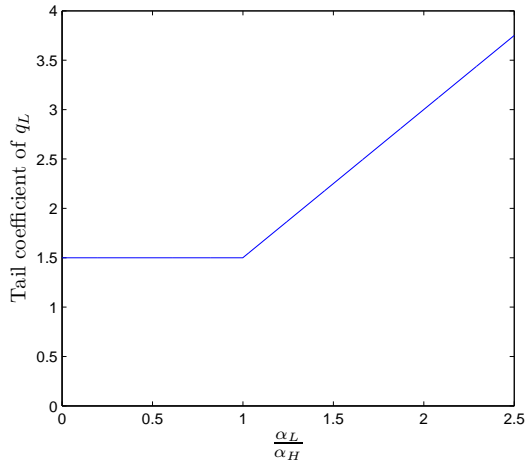}
\caption{The tail coefficient of $q_L$ under max-weight $\alpha$ scheduling, as a function of $\alpha_L/\alpha_H,$ for $C_H=2.5.$}\
\label{TCplot}
\end{center}
\end{figure}
In Figure~\ref{TCplot}, we show the tail coefficient of $q_L$ as a function of the ratio $\alpha_L/\alpha_H.$ We see that the tail coefficient is constant at the value $C_H-1$ as $\alpha_L/\alpha_H$ varies from 0 to 1. Recall that $\alpha_L/\alpha_H=1$ corresponds to max-weight scheduling, while $\alpha_L/\alpha_H\downarrow0$ corresponds to priority for $H$. Thus, the tail coefficient of $q_L$ under max-weight scheduling is the same as the tail coefficient under priority for $H$, implying that the max-weight policy leads to the\emph{ worst possible} asymptotic behavior for the light queue among all non-idling policies. However, the tail coefficient of $q_L$ begins to improve in proportion to the ratio $\alpha_L/\alpha_H$ in the regime where the light queue is given more importance.
\begin{remark}
If the heavy-tailed input has infinite variance ($C_H<2$), then it follows from Theorem~\ref{tailcoeff} that the expected delay in the light queue is infinite under max-weight scheduling. Thus, \cite[Proposition 5]{MMT09} is a special case of the above theorem.
\end{remark}
\section{Log-Max-Weight Scheduling}\label{logmax}

We showed in Theorem~\ref{tailcoeff} that the light queue occupancy distribution is necessarily heavy-tailed with a finite tail coefficient, under max-weight-$\alpha$ scheduling. On the other hand, the priority for $L$ policy which ensures the best possible asymptotic behavior for both queues, suffers from possible instability effects in more general queueing networks.

In this section, we propose and analyze the log-max-weight (LMW) policy. We show that the\emph{ light queue distribution is light-tailed} under LMW scheduling, i.e., that $\pr{q_L>b}$ decays exponentially fast in $b.$ However, unlike the priority for $L$ policy, LMW scheduling is throughput optimal even in more general settings. For our simple system model, we define the LMW policy as follows:

 In each time slot $t,$ the log-max-weight policy compares $$q_L(t)\gtreqless\log(1+q_H(t)),$$ and serves one packet from the queue that wins the comparison. Ties are broken in favor of the light queue. 
 
 The main idea in the LMW policy is to give preference to the light queue to a far greater extent than any max-weight-$\alpha$ policy. Specifically, for $\alpha_L/\alpha_H>1,$ the max-weight-$\alpha$ policy compares $q_L$ to a power of $q_H$ that is smaller than 1. On the other hand, LMW scheduling compares $q_L$ to a logarithmic function of $q_H,$ leading to a significant preference for the light queue. It turns out that this significant de-emphasis of the heavy queue with respect to the light queue is sufficient to ensure an exponential  decay for the distribution of $q_L$ in our setting.

 Furthermore, the LMW policy has another useful property when the heavy queue gets overwhelmingly large. Although the LMW policy significantly de-emphasizes the heavy queue, it does not \emph{ignore} it, unlike the priority for $L$ policy. That is, if the $H$ queue occupancy gets overwhelmingly large compared to $L$, the LMW policy will serve the $H$ queue. In contrast, the priority for $L$ policy will ignore any build-up in $H$, as long as $L$ is non-empty. This property turns out to be crucial in more complex queueing models, where throughput optimality is non-trivial to obtain. For example, when the queues have time-varying connectivity, the LMW policy will stabilize both queues for all rates within the rate region, whereas priority for $L$ leads to instability effects in $H$.

 Our main result in this section shows that under the LMW policy, $\pr{q_L>b}$ decays exponentially fast in $b,$ unlike under max-weight-$\alpha$ scheduling.

\begin{theorem}
\label{ubthm}Under log-max-weight scheduling, $q_L$ is light-tailed. Specifically, it holds that \begin{equation}\label{ub}\liminf_{b\to\infty}-\frac1b\log\pr{q_L\ge b}\geq\min(E_L,\ C_H-1),\end{equation} where $E_L$ is the intrinsic exponent, given by (\ref{innate}), (\ref{innate2}).
\end{theorem}

\emph{Proof:} Fix a small $\delta>0.$ We first write the equality

\begin{eqnarray}\pr{q_L\ge b}&=&\underbrace{\pr{q_L\ge b;\ \log(1+q_H)<\delta b}}_{(i)}\nonumber\\&+&\underbrace{\pr{q_L\ge b;\ (1-\delta)b\ge \log(1+q_H)\ge\delta b}}_{(ii)}\nonumber\\&+&\underbrace{\pr{q_L\ge b;\ \log(1+q_H)>(1-\delta)b}}_{(iii)}\label{threesplit}\end{eqnarray}
We will next upper bound each of the above three terms on the right.

\begin{itemize}

\item[(i)] $\pr{q_L\ge b;\ \log(1+q_H)<\delta b}:$ Intuitively, this event corresponds to an overflow of the light queue, when the light queue is not `exponentially large' in $b,$ i.e., $q_H<\exp(\delta b)-1.$ Suppose without loss of generality that this event happens at time $0.$ Denote by $-\tau\leq0$ the last instant when the heavy queue received service. Since $H$ has not received service since $-\tau$, it is clear that $\log(1+q_H(-\tau))<\delta b.$ Thus, $q_L(-\tau)<\delta b.$

    In the time interval $[-\tau+1,0]$ the light queue receives service in each slot. In spite of receiving all the service, it grows from less than $\delta b$ to overflow at time $0.$ This implies that every time the event in (i) occurs, there necessarily exists  $-u\le 0$ satisfying $$\sum_{i=-u+1}^0 (L(i)-1)>(1-\delta)b.$$ Therefore,

    $$\pr{q_L\ge b;\ \log(1+q_H)<\delta b}\le \pr{\exists u\geq 0\left|\sum_{i=-u+1}^0 (L(i)-1)>(1-\delta)b\right.}.$$ Letting $S_u=\sum_{i=-u+1}^0 L(i),$ the above inequality can be written as

    \begin{equation}\label{rw1}\pr{q_L\ge b;\ \log(1+q_H)<\delta b}\le \pr{\sup_{u\geq0}\left(S_u-u\right)>(1-\delta)b}.\end{equation}
    The right hand side of (\ref{rw1}) is precisely the probability of a single server queue fed by the process $L(\cdot)$ reaching the level $(1-\delta)b.$ Standard large deviation bounds are known for such an event. Specifically, from \cite[Lemma 1.5]{bigqueues}, we get
    $$\liminf_{b\to\infty}-\frac1b\log\pr{\sup_{u\geq0}S_u-u>(1-\delta)b}\geq\inf_{u>0}u\Lambda_L^*\left(1+\frac{1-\delta}u\right)$$\begin{equation}\label{lemma1.5} =\inf_{a>0}\frac{1-\delta}a\Lambda_L^*(1+a)=(1-\delta)E_L.
    \end{equation}
    From (\ref{rw1}) and (\ref{lemma1.5}), we see that for every $\epsilon>0$ and for large enough $b,$
\begin{equation}\label{ub1}\pr{q_L\ge b;\ \log(1+q_H)<\delta b}< \kappa_1e^{-b(1-\delta)(E_L-\epsilon)}.\end{equation}
\item[(iii)] Let us deal with the term (iii) before (ii). This is the regime where the overflow of $L$ occurs, along with $H$ becoming exponentially large in $b.$ We have
     \bearno\pr{q_L\ge b;\ \log(1+q_H)>(1-\delta)b}&=&\pr{q_L\ge b;\ q_H>e^{(1-\delta)b}-1}\\&\le&\pr{ q_L+q_H>e^{(1-\delta)b}}\enarno

    We have shown earlier in the proof of Theorem~\ref{ub_1} that for any non-idling policy,
$$\pr{q_L+q_H>M}<\kappa_2 M^{-(C_H-1-\epsilon)}$$ for every $\epsilon>0$ and some $\kappa_2>0.$ Therefore,
\begin{equation}\label{ub2}\pr{q_L\ge b;\ \log(1+q_H)>(1-\delta)b}<\kappa_2\exp\left(-(1-\delta)b(C_H-1-\epsilon)\right),\ \forall\ \epsilon>0.\end{equation}% Thus, \liminf_{b\to\infty}-\frac1b\log\pr{q_L\ge b;\ \log(1+q_H)>(1-\delta)b}\geq(C_H-1-\epsilon)(1-\delta),\ \forall\ \delta>0.
\item[(ii)] Let us now deal with the second term, $\pr{q_L\ge b;\ (1-\delta)b\ge \log(1+q_H)\ge\delta b}.$ Let us call this event $\mathcal E_2.$ Suppose this event occurs at time 0. Denote by $-\tau\leq0$ the last time during the current busy period that $H$ received service, and define $$\eta={\log(1+q_H(-\tau))}.$$ If $H$ never received service during the current busy period, we take $\tau$ to be equal to the last instant that the system was empty, and $\eta=0.$ We can deduce that $\eta\le (1-\delta)b,$ because $H$ receives no service in $[-\tau+1,0]$. It is also clear that $q_L(-\tau)<\eta .$ Therefore, $L$ grows from less than $\eta $ to more than $b,$ in spite of receiving all the service in $[-\tau+1,0].$ Using $u$ and $\xi$ as `dummy' variables that represent the possible values taken by $\tau$ and $\eta$ respectively, we can write
     \begin{eqnarray*}\pr{\mathcal E_2}&\le& \pr{\exists\ \xi\leq(1-\delta)b, u\geq0\left| S_u-u>b-\xi;\ q_H(-u)+q_L(-u)\geq e^{\xi}\right.}\\
     &\le&\sum_{\xi=0}^{(1-\delta)b}\pr{\exists\ u\geq0\left| S_u-u>b-\xi;\ q_H(-u)+q_L(-u)\geq e^{\xi }\right.}\\
     &\le&\sum_{\xi=0}^{(1-\delta)b}\sum_{u\geq0}\pr{S_u-u>b-\xi;\ q_H(-u)+q_L(-u)\geq e^{\xi }}\end{eqnarray*} where the last two steps are by the union bound. Notice now that for every $u\geq 0,$ the event $S_u-u>b-\xi$ is independent of the value of $q_H(-u)+q_L(-u),$ since these are determined by arrivals in disjoint intervals. Therefore, continuing from above
     \begin{eqnarray}&=&\sum_{\xi=0}^{(1-\delta)b}\sum_{u\geq0}\pr{S_u-u>b-\xi}\pr{q_H(-u)+q_L(-u)\geq e^{\xi }}\nonumber\\&\le&\sum_{\xi=0}^{(1-\delta)b}\sum_{u\geq0}\pr{S_u-u>b-\xi}\kappa_2 e^{-(C_H-1-\epsilon)\xi},\ \forall \ \epsilon>0\label{fromHub}\\ &\le&\sum_{\xi=0}^{(1-\delta)b}\kappa_1e^{-(E_L-\epsilon)(b-\xi)}\kappa_2 e^{-(C_H-1-\epsilon)\xi},\ \forall \ \epsilon>0.\label{bigq}
     \end{eqnarray}
    Equation (\ref{fromHub}) follows from (\ref{tailub_1}), and (\ref{bigq}) is a classical large deviation bound that follows, for example, from \cite[Lemma 1.5]{bigqueues}. Thus, for every $\epsilon>0,$
    \begin{equation}\pr{\mathcal E_2}\le \sum_{\xi=0}^{(1-\delta)b}\kappa_1\kappa_2 e^{-\left[(C_H-1-\epsilon)\xi + (E_L-\epsilon)(b-\xi)\right]}.\label{convexcomb}\end{equation} Let us now distinguish two cases:

    \begin{itemize}
    \item $C_H-1>E_L:$ In this case, we can bound the above probability as \begin{equation}\label{c>e1}\pr{\mathcal E_2}\le\kappa e^{-b(E_L-\epsilon)},\ \forall \epsilon>0,\end{equation} where $\kappa>0$ is some constant.

        \item $C_H-1\leq E_L:$ In this case, \begin{equation}\label{c<e1}\pr{\mathcal E_2}\le\kappa e^{-b(C_H-1-\epsilon)(1-\delta)},\ \forall \epsilon>0.\end{equation}
    \end{itemize}
\end{itemize}

Let us now put together the bounds on terms (i), (ii) and (iii) into Equation (\ref{threesplit}).

\begin{enumerate}

\item If $C_H-1>E_L,$ we get from (\ref{ub1}), (\ref{ub2}), and (\ref{c>e1}),
\begin{equation}\label{c>e2} \pr{q_L\ge b}<e^{-b(1-\delta)(E_L-\epsilon)}\left[\kappa_1+\kappa_2e^{-\left((1-\delta)b(C_H-1-E_L)\right)}+\kappa \right],
\end{equation}
from which it is immediate that
$$\liminf_{b\to\infty}-\frac1b\log\pr{q_L\ge b}\geq(1-\delta)(E_L-\epsilon).$$ Since the above is true for each $\epsilon$ and $\delta,$ we get
\begin{equation}\label{finalub1}\liminf_{b\to\infty}-\frac1b\log\pr{q_L\ge b}\geq E_L.\end{equation}

\item If $C_H-1\leq E_L,$ we get from (\ref{ub1}), (\ref{ub2}), and (\ref{c<e1}),
\begin{equation}\label{c<e2} \pr{q_L\ge b}<e^{-b(1-\delta)(C_H-1-\epsilon)}\left[\kappa_1e^{-\left((1-\delta)b(E_L-C_H+1)\right)}+\kappa_2+\kappa \right],
\end{equation}
from which it is immediate that
$$\liminf_{b\to\infty}-\frac1b\log\pr{q_L\ge b}\geq(1-\delta)(C_H-1-\epsilon).$$
Since the above is true for each $\epsilon$ and $\delta,$ we get
\begin{equation}\label{finalub2}\liminf_{b\to\infty}-\frac1b\log\pr{q_L\ge b}\geq C_H-1.\end{equation}
\end{enumerate}
Theorem~\ref{ubthm} now follows from (\ref{finalub1}) and (\ref{finalub2}).\qed

Thus, the light queue tail is upper bounded by an exponential term, whose rate of decay is given by the smaller of the intrinsic exponent $E_L,$ and $C_H-1.$ We remark that Theorem~\ref{ubthm} utilizes only the light-tailed nature of $L(\cdot),$ and the tail coefficient of $H(\cdot).$ Specifically, we do not need to assume any regularity property such as $H(\cdot)\in\mathcal{OR}$ for the result to hold. However, if we assume that the tail of $H(\cdot)$ is regularly varying, we can obtain a matching lower bound to the upper bound in Theorem~\ref{ubthm}.
\begin{theorem}\label{regvar}
Suppose that $H(\cdot)\in\mathcal R (C_H).$ Then, under LMW scheduling, the tail distribution of $q_L$ satisfies an LDP with rate function given by $$\lim_{b\to\infty}-\frac1b\log\pr{q_L\ge
b}=\min(E_L,\ C_H-1).$$
\end{theorem}
\emph{Proof:} In light of Theorem~\ref{ubthm}, it is enough to
prove that $$\limsup_{b\to\infty}-\frac1b\log\pr{q_L\ge
b}\le\min(E_L,\ C_H-1).$$

Let us denote by $q_L^{(p)}$ the queue length of the light queue,
when it is given complete priority over $H.$ Note that
$\pr{q_L^{(p)}>b}$ is a lower bound on the overflow probability
under \emph{any} policy, including LMW. Therefore, for all $b>0,$
$\pr{q_L\ge b}\geq\pr{q_L^{(p)}>b}.$ This implies

\begin{eqnarray}\label{lb1}\limsup_{b\to\infty}-\frac1b\log\pr{q_L\ge b}&\le&\limsup_{b\to\infty}-\frac1b\log\pr{q_L^{(p)}>b}= E_L,\end{eqnarray}  where the last step is from (\ref{LDP}).

Next, we can show, following the arguments in  Proposition~\ref{fict} and
Theorem~\ref{fictlb} that $$\pr{q_L\ge b}\geq \lambda_H\pr{H_R\geq e^b-1;\ \sum_{i=1}^{H_A}L(i)\ge b}.$$ But arguing similarly to Lemma~\ref{mainlemma}, we can show that $$\pr{H_R\geq e^b-1;\ \sum_{i=1}^{H_A}L(i)\ge b}\sim\pr{H_R\geq e^b-1}.$$ Thus,$$\pr{q_L\ge b}\gtrsim \pr{H_R\geq
e^b-1}.$$ Next, since $H(\cdot)$ is regularly varying with tail coefficient
$C_H,$ $H_R$ is also regularly varying with tail coefficient $C_H-1,$ so that  $\pr{H_R\geq
e^b-1}=U(e^b)e^{-b(C_H-1)}.$ Finally we can write
$$\limsup_{b\to\infty}-\frac1b\log\pr{q_L\ge
b}\le \limsup_{b\to\infty}-\frac1b\log\pr{H_R\geq
e^b-1}=C_H-1-\limsup_{b\to\infty}\frac{\log U(e^b)}b.$$ The final limit supremum is shown to be zero in Lemma~\ref{slowlemma}, using a representation theorem for slowly varying functions. Thus,
\begin{eqnarray}\label{lb2}\limsup_{b\to\infty}-\frac1b\log\pr{q_L\ge b}\leq C_H-1.\end{eqnarray}
Equations (\ref{lb1}) and (\ref{lb2}) imply the theorem. \qed
\begin{figure}
\begin{center}
\includegraphics[width=8.5cm]{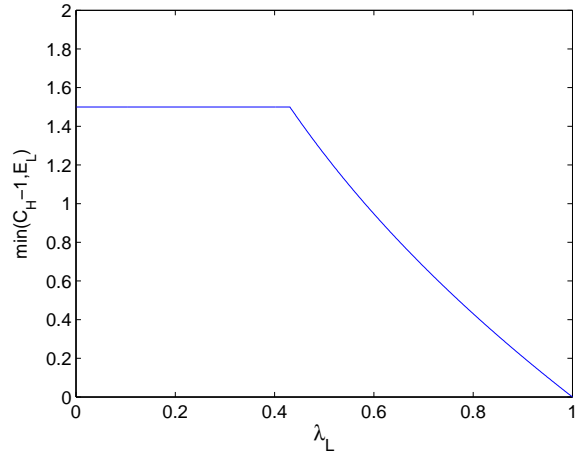}
\caption{The large deviation exponent for $q_L$ under LMW scheduling, as a function of $\lambda_L.$ The light queue is fed by Poisson bursts, and $C_H=2.5.$}
\label{ld}
\end{center}
\end{figure}

Figure~\ref{ld} shows the large deviation exponent given by Theorem~\ref{regvar} as a function of $\lambda_L$, for $C_H=2.5,$ and Poisson inputs feeding the light queue. There are two distinct regimes in the plot, corresponding to two fundamentally different modes of overflow. For relatively large values of $\lambda_L,$ the exponent for the LMW policy equals $E_L,$ the intrinsic exponent. In this regime, the light queue overflows entirely due to atypical behavior in the input process $L(\cdot).$ In other words, $q_L$ would have grown close to the level $b$ even if the heavy queue was absent. This mode of overflow is more likely for larger values of $\lambda_L,$ which explains the diminishing exponent in this regime.

The flat portion of the curve in Figure~\ref{ld} corresponds to a second overflow mode. In this regime, the overflow of the light queue occurs due to extreme misbehavior on the part of the heavy-tailed input. Specifically, the heavy queue becomes larger than $e^b$ after receiving a very large burst. After this instant, the heavy queue hogs all the service, and the light queue gets starved until it gradually builds up to the level $b.$ In this regime, the light queue input behaves typically, and plays no role in the overflow of $L$. That is, the exponent is independent of $\lambda_L,$ being equal to a constant $C_H-1.$ The exponent is decided entirely by the `burstiness' of the heavy-tailed traffic, which is reflected in the tail coefficient.

\section{Concluding Remarks}\label{concl}
We considered a system of parallel queues fed by a mix of heavy-tailed and light-tailed traffic, and served by a single server. We studied the asymptotic behavior of the queue size distributions under various scheduling policies. We showed that the occupancy distribution of the heavy queue is asymptotically insensitive to the scheduling policy used, and inevitably heavy-tailed. In contrast, the light queue occupancy distribution can be heavy-tailed or light-tailed depending on the scheduling policy.

The major contribution of the paper is in the derivation of an exact asymptotic characterization of the light queue occupancy distribution, under max-weight-$\alpha$ scheduling. We showed that the light queue distribution is heavy-tailed with a finite tail coefficient under max-weight-$\alpha$ scheduling, for any values of the scheduling parameters. However, the tail coefficient can be improved by choosing the scheduling parameters to favor the light queue. We also observed that `plain' max-weight scheduling leads to the worst possible asymptotic behavior of the light queue distribution, among all non-idling policies.

Another important contribution of the paper is the log-max-weight policy, and the corresponding asymptotic analysis. We showed that the light queue occupancy distribution is light-tailed under LMW scheduling, and explicitly derived an exponentially decaying upper bound on the tail of the light queue distribution. Additionally, the LMW policy also has the desirable property of being throughput optimal in a general queueing network.

Although we study a very simple queueing network in this paper, we believe that the insights obtained from this study are valuable in much more general settings. For instance, in a general queueing network with a mix of light-tailed and heavy-tailed traffic flows, we expect that the celebrated max-weight policy has the tendency to `infect' competing light-tailed flows  with heavy-tailed asymptotics. A similar effect was also noted in \cite{MMT09}, in the context of expected delay.

We also believe that the LMW policy occupies a unique `sweet spot' in the context of scheduling light-tailed traffic in the presence of heavy-tailed traffic. This is because the LMW policy de-emphasizes the heavy-tailed flow sufficiently to maintain good light queue asymptotics, while also ensuring network-wide stability.

For future work, we propose the extension of the results in this paper to more general single-hop and multi-hop network models. Even in the context of parallel queues, the incorporation of time-varying channel models presents an interesting direction.
\appendix
\section{Technical Lemmas}
\begin{lemma}
\label{joint2}
$\pr{H_R\geq m,\ H_A\geq n}=\pr{H_R\geq m+n}$
\end{lemma}
\emph{Proof:} Using (\ref{joint}) and (\ref{residue}),
\begin{eqnarray*}\pr{H_R\geq m,\ H_A\geq n}&=&\sum_{k\ge m}\sum_{l\ge n}\frac{\pr{H_+=k+l}}{\expect{H_+}}\\
&=&\sum_{k\ge m}\sum_{p=k+n}^{\infty}\frac{\pr{H_+=p}}{\expect{H_+}}\\
&=&\sum_{k\ge m}\pr{H_R=k+n}\\
&=&\pr{H_R\geq m+n}.
\end{eqnarray*}\qed
%
%\begin{lemma}\label{ERlemma}
%If $H(\cdot)\in\mathcal{OR},$ then $H_R\in\mathcal{ER},$ and
%\begin{equation}\label{sup}
%\sup_n \frac{n\pr{H_+>n}}{\pr{H_R>n}}<\infty.
%\end{equation}
%\end{lemma}
%\emph{Proof:} See \cite[Lemma 4.2(i)]{cline94}.

\begin{lemma}\label{rand_sum}
Let $N\in\mathcal{IR}$ be a non-negative integer valued random variable. Let
$X_i, i\geq1$ be i.i.d. non-negative light-tailed random variables, with mean $\mu,$ independent of $N.$ Define
$$S_N=\sum_{i=1}^N X_i.$$ Then, \[\pr{S_N>b}\sim\pr{N>b/\mu}.\]
\end{lemma}
\emph{Proof:} For notational ease, we will prove the result for
$\mu=1,$ although the result and proof technique are applicable for
any $\mu>0.$ First, for a fixed $\delta>0,$ we have

\begin{eqnarray}\label{ub_rand}
\pr{S_N>b}&=&\pr{S_N>b;\ N\le b(1-\delta)}+\pr{S_N>b;\
N>b(1-\delta)}\nonumber\\&<&\pr{S_{\lfloor
b(1-\delta)\rfloor}>b}+\pr{N>b(1-\delta)}.\end{eqnarray} Next, we
write a lower bound:

\begin{eqnarray}\label{lb_rand}
\pr{S_N>b}&\ge& \pr{S_N>b;\ N>b(1+\delta)}\nonumber\\
&=&\pr{N>b(1+\delta)}-\pr{S_N\leq b;\ N>b(1+\delta)}\nonumber\\
&\ge&\pr{N>b(1+\delta)}-\pr{S_{\lceil b(1+\delta)\rceil}\leq b}.
\end{eqnarray}

Since the $X_i$ have a well defined moment generating function,
their sample average satisfies an exponential concentration
inequality around the mean. Specifically, we can show using the
Chernoff bound that there exist constants $\kappa,\eta$ such that
\[ \pr{S_{\lfloor b(1-\delta)\rfloor}>b}<\kappa e^{-b\eta}.\] Thus,
it follows that \begin{equation}\label{o1}\pr{S_{\lfloor
b(1-\delta)\rfloor}>b}=o(\pr{N>b})\end{equation} as $b\to\infty.$
Similarly,\begin{equation}\label{o2}\pr{S_{\lfloor
b(1+\delta)\rfloor}\leq b}=o(\pr{N>b}).\end{equation}

Next, getting back to (\ref{ub_rand}),

\[\limsup_{b\to\infty}\frac{\pr{S_N>b}}{\pr{N>b}}\leq\limsup_{b\to\infty}\frac{\pr{S_{\lfloor
b(1-\delta)\rfloor}>b}}{\pr{N>b}}+\limsup_{b\to\infty}\frac{\pr{N>b(1-\delta)}}{\pr{N>b}}.\]
The first term on the right hand side is zero in view of (\ref{o1}),
so that for all $\delta,$ we have
\[\limsup_{b\to\infty}\frac{\pr{S_N>b}}{\pr{N>b}}\le\limsup_{b\to\infty}\frac{\pr{N>b(1-\delta)}}{\pr{N>b}}.\]
Taking the limit as $\delta\downarrow 0, $
\begin{equation}\label{lsup}\limsup_{b\to\infty}\frac{\pr{S_N>b}}{\pr{N>b}}\le\lim_{\delta\downarrow0}\limsup_{b\to\infty}\frac{\pr{N>b(1-\delta)}}{\pr{N>b}}=1\end{equation}
The final limit is unity, by the definition of the class $\mathcal{IR}$.
Similarly, we can show using (\ref{lb_rand}), (\ref{o2}) and the
intermediate-regular variation of the tail of $N$ that
\begin{equation}\label{linf}\liminf_{b\to\infty}\frac{\pr{S_N>b}}{\pr{N>b}}\ge
1.\end{equation} Equations (\ref{lsup}) and (\ref{linf}) imply the
result. \qed

The above lemma can be proved under more general assumptions than stated here, see \cite{RS2008}.
\begin{lemma}
\label{mainlemma} If $H(\cdot)\in\mathcal{OR},$ we have
\begin{equation}\pr{H_R\ge b^{\alpha_L/\alpha_H},\ \sum_{i=1}^{H_A}L(i)\ge b}\sim\left\{\begin{array}{cc} \pr{H_R\ge \frac{b}{\lambda_L}}&\frac{\alpha_L}{\alpha_H}<1,\\ \pr{H_R\ge b+\frac{b}{\lambda_L}}&\frac{\alpha_L}{\alpha_H}=1,\\ \pr{H_R\ge b^{\alpha_L/\alpha_H}}&\frac{\alpha_L}{\alpha_H}>1.\end{array}\right.
\end{equation}
\end{lemma}
\emph{Proof:} In this proof, let us take $\lambda_L=1$ for notational simplicity, although the same proof technique works without this assumption. Denote $S_n=\sum_{i=1}^n L(i).$ We first get an upper bound. For
every $\delta>0$, we have
\begin{eqnarray}
\pr{H_R\ge b^{\alpha_L/\alpha_H};\ S_{H_A}\ge b}&=&\nonumber\\\pr{H_R\ge b^{\alpha_L/\alpha_H};\ S_{H_A}\ge b;\ H_A<b(1-\delta)}&+&\pr{H_R\ge b^{\alpha_L/\alpha_H};\ S_{H_A}\ge b;\ H_A>b(1-\delta)}\nonumber\\<\pr{S_{H_A}\ge b;\ H_A<b(1-\delta)}&+&\pr{H_R\ge b^{\alpha_L/\alpha_H};\ H_A>b(1-\delta)}\\ \le \pr{S_{H_A}\ge b;\ H_A<b(1-\delta)}&+&\pr{H_R\ge b^{\alpha_L/\alpha_H}+b(1-\delta)}\label{indep}\\<\pr{S_{\lfloor b(1-\delta)\rfloor}>b}&+&\pr{H_R\ge b^{\alpha_L/\alpha_H}+b(1-\delta)}\label{lem}.
%\pr{H_R\ge b^{\alpha_L/\alpha_H};\ S_{H_A}\ge b}&=&\nonumber\\\pr{\cdot\ ;\ H_A<b(1-\delta)}&+&\pr{\cdot\ ;\ H_A\geq b(1-\delta)}\nonumber\\ \leq \pr{S_{H_A}\ge b;\ H_A<b(1-\delta)}&+&\pr{H_R\ge b^{\alpha_L/\alpha_H};\ H_A\geq b(1-\delta)}\nonumber\\ \le \pr{S_{H_A}\ge b;\ H_A<b(1-\delta)}&+&\pr{H_R\ge b^{\alpha_L/\alpha_H}+b(1-\delta)}\label{indep}\\<\pr{S_{\lfloor b(1-\delta)\rfloor}>b}&+&\pr{H_R\ge b^{\alpha_L/\alpha_H}+b(1-\delta)}\label{lem}.
\end{eqnarray}
In (\ref{indep}) we have utilized Lemma~\ref{joint2}, and in Equation (\ref{lem}), we have used the independence of $H_A$ and $L(\cdot).$ Next, let us derive a lower bound.
\begin{eqnarray}
\pr{H_R\ge b^{\alpha_L/\alpha_H};\ S_{H_A}\ge b}&\ge& \pr{H_R\ge b^{\alpha_L/\alpha_H};\ S_{H_A}\ge b; \ H_A>b(1+\delta)}=\nonumber\\
\pr{H_R\ge b^{\alpha_L/\alpha_H};\ H_A>b(1+\delta)}&-&\pr{H_R\ge b^{\alpha_L/\alpha_H};\ S_{H_A}< b; \ H_A>b(1+\delta)}\ge\nonumber\\
\pr{H_R\ge b^{\alpha_L/\alpha_H};\ H_A>b(1+\delta)}&-&\pr{S_{H_A}< b; \ H_A>b(1+\delta)}\geq\nonumber\\
\pr{H_R\ge b^{\alpha_L/\alpha_H}+b(1+\delta)}&-&\pr{S_{\lceil b(1+\delta)\rceil}\leq b}.\label{lb_lem}
\end{eqnarray}
Equation (\ref{lb_lem}) uses Lemma~\ref{joint2}. Now, observe that the terms  $\pr{S_{\lfloor b(1-\delta)\rfloor}>b}$ in (\ref{lem}) and $\pr{S_{\lceil b(1+\delta)\rceil}\leq b}$ in (\ref{lb_lem}) decay exponentially fast as $b\to\infty,$ for any $\delta>0.$ This is because $L(\cdot)$ is light-tailed, and their sample average satisfies an exponential concentration inequality around the mean (unity). More precisely, a Chernoff bound can be used to show that
\begin{equation}
\label{o3}
\pr{S_{\lfloor b(1-\delta)\rfloor}>b}=o\left(\pr{H_R\ge b^{\alpha_L/\alpha_H}+b}\right),
\end{equation}
and
\begin{equation}
\label{o4}
\pr{S_{\lceil b(1+\delta)\rceil}\leq b}=o\left(\pr{H_R\ge b^{\alpha_L/\alpha_H}+b}\right).
\end{equation}
%See \cite[Theorem 3.1]{RS2008} for a proof of (\ref{o1}) and (\ref{o2}) under more general assumptions. Let us now consider the three cases in the statement of the lemma separately.
\emph{Case (i): $\frac{\alpha_L}{\alpha_H}<1.$}
Using (\ref{lem}), we write
$$\limsup_{b\to\infty}\frac{\pr{H_R\ge b^{\alpha_L/\alpha_H};\ S_{H_A}\ge b}}{\pr{H_R\ge b}}\le\limsup_{b\to\infty}\frac{\pr{S_{\lfloor b(1-\delta)\rfloor}>b}}{\pr{H_R\ge b}}+\limsup_{b\to\infty}\frac{\pr{H_R\ge b^{\alpha_L/\alpha_H}+b(1-\delta)}}{\pr{H_R\ge b}}.$$ The first limit supremum on the right is zero in view of (\ref{o3}). Since $\frac{\alpha_L}{\alpha_H}<1,$ we can write
$$\limsup_{b\to\infty}\frac{\pr{H_R\ge b^{\alpha_L/\alpha_H};\ S_{H_A}\ge b}}{\pr{H_R\ge b}}\le\limsup_{b\to\infty}\frac{\pr{H_R\ge b(1-\delta)}}{\pr{H_R\ge b}}, \ \forall \delta>0.$$ Thus, \begin{equation}\label{sup1}\limsup_{b\to\infty}\frac{\pr{H_R\ge b^{\alpha_L/\alpha_H};\ S_{H_A}\ge b}}{\pr{H_R\ge b}}\le \lim_{\delta\downarrow0}\limsup_{b\to\infty}\frac{\pr{H_R\ge b(1-\delta)}}{\pr{H_R\ge b}}=1.\end{equation}
The final limit is unity, because according to Lemma~\ref{ERlemma}, $H(\cdot)\in\mathcal{OR}$ implies $H_R\in\mathcal{ER}.$ Since $\mathcal{ER}\subset\mathcal{IR},$ the final limit in (\ref{sup1}) is unity, by the definition of intermediate-regular variation (Definition~\ref{reg}).

Along similar lines, we can use (\ref{lb_lem}), (\ref{o4}), and the fact that $H_R\in\mathcal{IR}$ to show that
\begin{equation}\label{inf1}\liminf_{b\to\infty}\frac{\pr{H_R\ge b^{\alpha_L/\alpha_H};\ S_{H_A}\ge b}}{\pr{H_R\ge b}}\ge \lim_{\delta\downarrow0}\liminf_{b\to\infty}\frac{\pr{H_R\ge b(1+2\delta)}}{\pr{H_R\ge b}}=1.\end{equation}
Equations (\ref{sup1}) and (\ref{inf1}) imply that $${\pr{H_R\ge b^{\alpha_L/\alpha_H};\ S_{H_A}\ge b}}\sim \pr{H_R\ge b},$$ which implies Lemma~\ref{mainlemma} for $\frac{\alpha_L}{\alpha_H}<1,$ and $\lambda_L=1.$
\\\emph{Case (ii): $\frac{\alpha_L}{\alpha_H}=1.$}
Similar to the previous case. Here, we get $${\pr{H_R\ge b;\ S_{H_A}\ge b}}\sim \pr{H_R\ge 2b}.$$
\emph{Case (iii): $\frac{\alpha_L}{\alpha_H}>1.$}

For the upper bound, we have from (\ref{lem}) and (\ref{o1}),
$$\limsup_{b\to\infty}\frac{\pr{H_R\ge b^{\alpha_L/\alpha_H};\ S_{H_A}\ge b}}{\pr{H_R\ge b^{\alpha_L/\alpha_H}}}\le\limsup_{b\to\infty}\frac{\pr{H_R\ge b^{\alpha_L/\alpha_H}+b(1-\delta)}}{\pr{H_R\ge b^{\alpha_L/\alpha_H}}}\leq 1.$$
Similarly, for the lower bound, we have from (\ref{lb_lem}) and (\ref{o2}),
$$\liminf_{b\to\infty}\frac{\pr{H_R\ge b^{\alpha_L/\alpha_H};\ S_{H_A}\ge b}}{\pr{H_R\ge b^{\alpha_L/\alpha_H}}}\geq \liminf_{b\to\infty}\frac{\pr{H_R\ge b^{\alpha_L/\alpha_H}+b(1+\delta)}}{\pr{H_R\ge b^{\alpha_L/\alpha_H}}},$$
$$\geq \liminf_{b\to\infty}\frac{\pr{H_R\ge b^{\alpha_L/\alpha_H}(1+\delta)}}{\pr{H_R\ge b^{\alpha_L/\alpha_H}}}, \ \forall \delta>0.$$ Thus,
$$\liminf_{b\to\infty}\frac{\pr{H_R\ge b^{\alpha_L/\alpha_H};\ S_{H_A}\ge b}}{\pr{H_R\ge b^{\alpha_L/\alpha_H}}}\geq \lim_{\delta\downarrow0}\liminf_{b\to\infty}\frac{\pr{H_R\ge b^{\alpha_L/\alpha_H}(1+\delta)}}{\pr{H_R\ge b^{\alpha_L/\alpha_H}}}=1,$$ where the last limit is unity due to the intermediate-regular variation of $H_R.$
Therefore, we can conclude for $\frac{\alpha_L}{\alpha_H}>1$ that $${\pr{H_R\ge b^{\alpha_L/\alpha_H};\ S_{H_A}\ge b}}\sim \pr{H_R\ge b^{\alpha_L/\alpha_H}}.$$ Lemma~\ref{mainlemma} is now proved. \qed

\begin{lemma}
\label{slowlemma}
For any slowly varying function $U(\cdot),$ $$\lim_{a\to\infty}\frac{\log U(a)}{\log a}=0.$$
\end{lemma}
\emph{Proof:} We use the representation theorem for slowly varying functions derived in \cite{GS73}. For every slowly varying function $U(\cdot),$ there exists a $B>0$ such that for all $x\geq B,$ the function can be written as $$U(x)=\exp\left(v(x)+\int_B^x\frac {\zeta(y)}y{\rm d}y\right),$$ where $v(x)$ converges to a finite constant, and $\zeta(x)\to 0$ as $x\to\infty.$ Therefore, $$\lim_{a\to\infty}\frac{\log U(a)}{\log a}=\lim_{a\to\infty}\frac {v(a)+\int_B^a\frac {\zeta(y)}y{\rm d}y}{\log a}=\lim_{a\to\infty}\frac {\int_B^a\frac {\zeta(y)}y{\rm d}y}{\log a},$$ where the last step is because $v(a)$ converges to a constant. Next, given any $\epsilon>0,$ choose $C(\epsilon)$ such that $\left|\zeta(a)\right|<\epsilon,\ \forall\ a>C(\epsilon).$ Then, we have $$\lim_{a\to\infty}\frac {\left|\int_B^a\frac {\zeta(y)}y{\rm d}y\right|}{\log a}\leq\lim_{a\to\infty}\frac {\int_B^{C(\epsilon)}\frac {\left|\zeta(y)\right|}y{\rm d}y+\int_{C(\epsilon)}^a\frac {\left|\zeta(y)\right|}y{\rm d}y}{\log a}\leq\lim_{a\to\infty}\frac{\epsilon\log \frac a{C(\epsilon)}}{\log a}=\epsilon.$$ Since the above is true for every $\epsilon,$ the result follows. \qed

\bibliographystyle{IEEEtranS}
\bibliography{mwa}

\end{document}